\documentclass[aip,reprint]{revtex4-2}
\usepackage[english]{babel}
\usepackage{graphicx}

\begin{document}
	\title{Nearly ideal memristive functionality based on viscous magnetization dynamics}
	
	\author{Sergei Ivanov$^1$}
	\author{Sergei Urazhdin$^1$}
	
	\affiliation{$^1$Department of Physics, Emory University, Atlanta, GA, USA.}

	\begin{abstract}
We experimentally demonstrate a proof-of-principle implementation of an almost ideal memristor - a two-terminal circuit element whose resistance is approximately proportional to the integral of the input signal over time. The demonstrated device is based on a thin-film ferromagnet/antiferromagnet bilayer, where magnetic frustration results in viscous magnetization dynamics enabling memristive functionality, while the external magnetic field plays the role of the driving input. The demonstrated memristor concept is amenable to downscaling and can be adapted for electronic driving, making it attractive for applications in neuromorphic circuits.\end{abstract}

	\maketitle

{\it Introduction.} Memristor - a two-terminal electronic device whose resistance is ideally proportional to the integral of the input signal, such as current or magnetic field - is one of the most promising candidates for the hardware implementation of synapses in artificial neural networks~\cite{handbook,nanoscale,ielmini}. According to the original definition~\cite{1083337,1454361}, an ideal memristor can be described by the equations
\begin{equation}
\label{eq:1}
\frac{dR}{dt}=aI(t), V(t)=R(t)I(t), R(t=0)=R_0,
\end{equation}
where $R$ is the resistance, $I(t)$ is the input signal, $V(t)$ is the output signal, $R_0$ and $a$ are constants. The input signal $I(t)$ is typically the current or voltage. In the experimental demonstration discussed below, we for simplicity utilize magnetic field as the input driving the memristor via Zeeman coupling. This approach is equivalent to current driving, since current can be converted to the magnetic field via Ampère's law. At the nanoscale, direct current driving can be implemented using spin torque~\cite{doi:10.1063/5.0018411}.

\begin{figure}
	\includegraphics[width=\columnwidth]{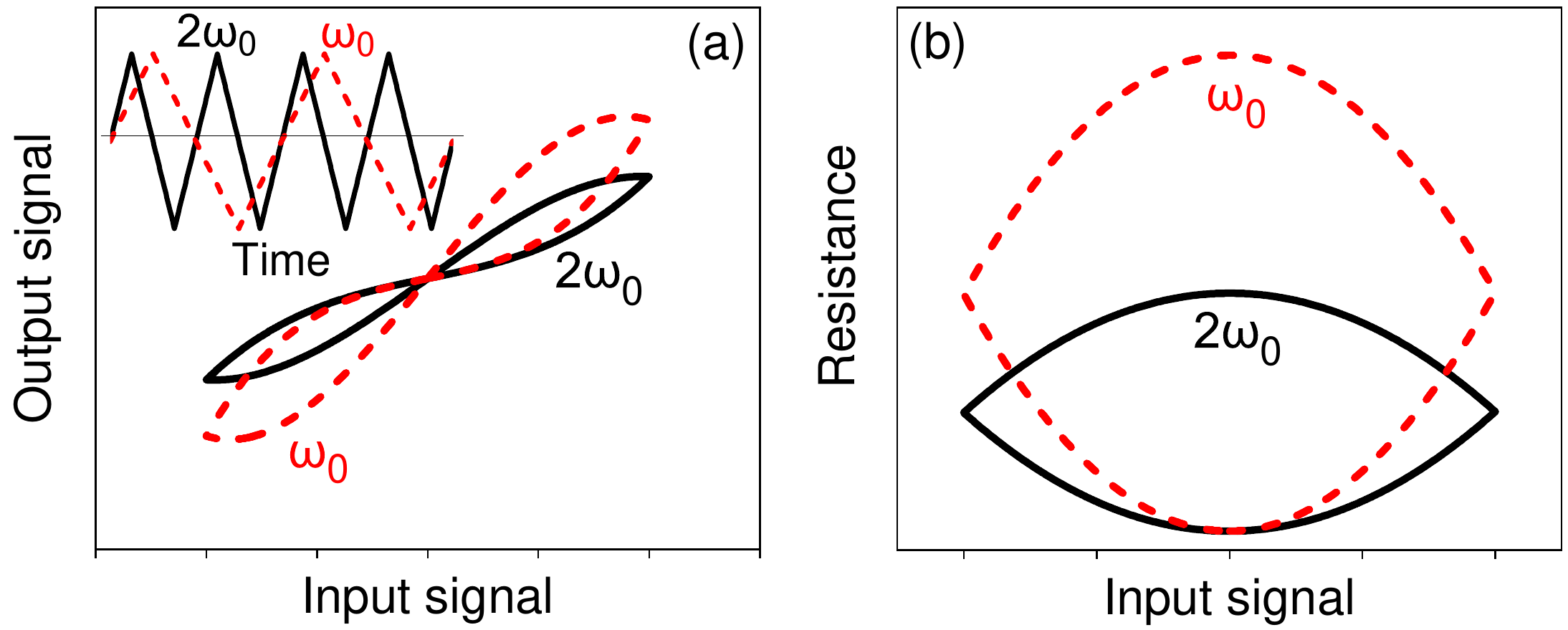}
	\caption{Characteristics of an ideal memristor. (a) Output  vs input signal, (b) Resistance vs input signal, for two different frequencies of the periodic triangular input (inset), as labeled.}
	\label{fig_1}
\end{figure}

For a periodic input signal (inset in Fig.~\ref{fig_1}), the solution of Eqs.~(\ref{eq:1}) is a pinched V-I hysteresis loop (Fig.~\ref{fig_1}(a)). The R-I loop is lens-shaped for the triangular input waveform (Fig.~\ref{fig_1}(b)).

In the integral form, Eq.~(\ref{eq:1}) is
\begin{equation}
\label{eq:2}
    R(t) = R_0 + a \int_{0}^{t} I(\tau)  d\tau.
\end{equation}

According to Eq.~(\ref{eq:2}), the area of the R-I hysteresis loop (Fig.~\ref{fig_1}(b)) increases linearly with the period of the input signal, or equivalently decreases inversely with the sweep rate $\omega$, as shown in Fig.~\ref{fig_1}(b) for two values of $\omega$. We note that Eq.~\ref{eq:2} is unphysical, as it implies that $R<0$ can be achieved for $aI<0$. However, it may be possible to achieve an approximately integral dependence of $R$ on the input over some range of the latter, in a device that can be defined as a nearly ideal memristor.  

Almost 50 years after the memristor was proposed,  functionality even approximately described by Eq.~(\ref{eq:1}) has not been achieved~\cite{Pershin_2018}. Instead, a variety of generalized memristors - devices that exhibit some dependence of resistance on the input history - have been developed~\cite{Wang2015}. These devices exhibit a pinched V-I hysteresis loop, but are not described by Eq.~(\ref{eq:1})~\cite{Pershin_2018,Kim_2020}. For instance, resistance switching memristors are characterized by a V-I hysteresis loop similar to that in Fig.~\ref{fig_1}(a), but  is  independent of the sweep rate $\omega$. Their resistance also does not vary at a constant input signal, as would be expected from Eq.~(\ref{eq:1}). These limitations restrict applications of such devices in neuromorphic hardware.

Here, we present a proof-of-principle experimental demonstration of nearly ideal memristive functionality approximately described by Eq.~(\ref{eq:1}).   
Our approach is based on the observation that the ideal memristive behavior in the differential form $\frac{dR}{dt}=aI(t)$ can be interpreted as a viscous response of $R$ to the input $I$. Here, we use the term "viscous" by analogy to Newton's law in the presence of viscous friction, $m\ddot{x}+\nu \dot{x}=F_{ext}(t)$, where $F_{ext}$ is the external force, and $\nu$ is viscosity. At small Reynolds numbers is reduced to $\dot{x}=F_{ext}(t)/\nu$, the same form as Eq.~(\ref{eq:1}) for the ideal memristor.

We expect that nearly ideal memristive functionality can be achieved using a variety of physical implementations of viscous dynamics of mechanical, electronic, or magnetic degrees of freedom. Since the latter do not require physical motion, they generally exhibit superior endurance, repeatability, and reproducibility, which have motivated extensive exploration of magnetic generalized memristors. The most common design is based on the domain wall (DW) pinned at various positions in a magnetic nanowire incorporated into a magnetic tunnel junction~\cite{4781542}. However, DW motion becomes increasingly discrete at nanoscale, and either requires an input  threshold or is randomized due to thermal fluctuations. These issues detrimental to memristive functionality can be described as deviations from viscous DW motion.

Another recent approach utilizes a dry friction-like effect of magnetization pinning in a nanoscale tunnel junction
by the high density of pinning centers in ferromagnets (Fs) sandwiched with thin-film antiferromagnets (AFs)~\cite{PhysRevApplied.12.044029,9108129}. This approach is amenable to downscaling, but unlike viscous friction, dry friction entails a significant input threshold for the dynamical response and does not result in ideal memristive behaviors.

F/AF systems have been extensively studied in the context of  exchange bias (EB) - unidirectional asymmetry of the magnetic hysteresis loop~\cite{PhysRev.105.904}. It is well-established that thick AF in F/AF bilayers form a multidomain state to accommodate frustrated exchange interaction at the F/AF interface~\cite{Radu2008}. Frustration in sufficiently thin AFs was predicted to result in the formation of a ''Heisenberg domain state" (HDS) that consists of tightly-packed AF DWs~\cite{PhysRevB.35.3679,PhysRevB.37.7673,doi:10.1063/1.340591}, but the implications for the dynamical magnetic properties remained virtually unexplored. 

In AF alloys such as IrMn utilized in Ref.~\cite{PhysRevApplied.12.044029}, the magnetization dynamics is likely dominated by pinning on magnetic defects associated with variations of local stoichiometry, resulting in dry friction-like behaviors. In contrast, time-domain and ac susceptibility  measurements in Fs sandwiched with high-quality AFs CoO and NiO demonstrated viscous AF dynamics close to EB blocking temperature $T_B$~\cite{PhysRevB.92.174416,PhysRevB.94.024422,PhysRevB.97.054402,URAZHDIN201975}. At lower temperatures, the viscosity was shown to rapidly increase, leading to magnetic freezing. Thus, HDS was identified as a correlated spin glass. Such a state can be particularly advantageous for memristive applications, since exchange interaction that underlies the frustration is large, provides a large-scale energy landscape that can stabilize a multitude of spin states even in nanoscale devices.

We build on these findings to demonstrate nearly ideal memristive functionality in the viscous state of a thin-film bilayer of Permalloy (Py) with NiO.

\begin{figure}
	\includegraphics[width=\columnwidth]{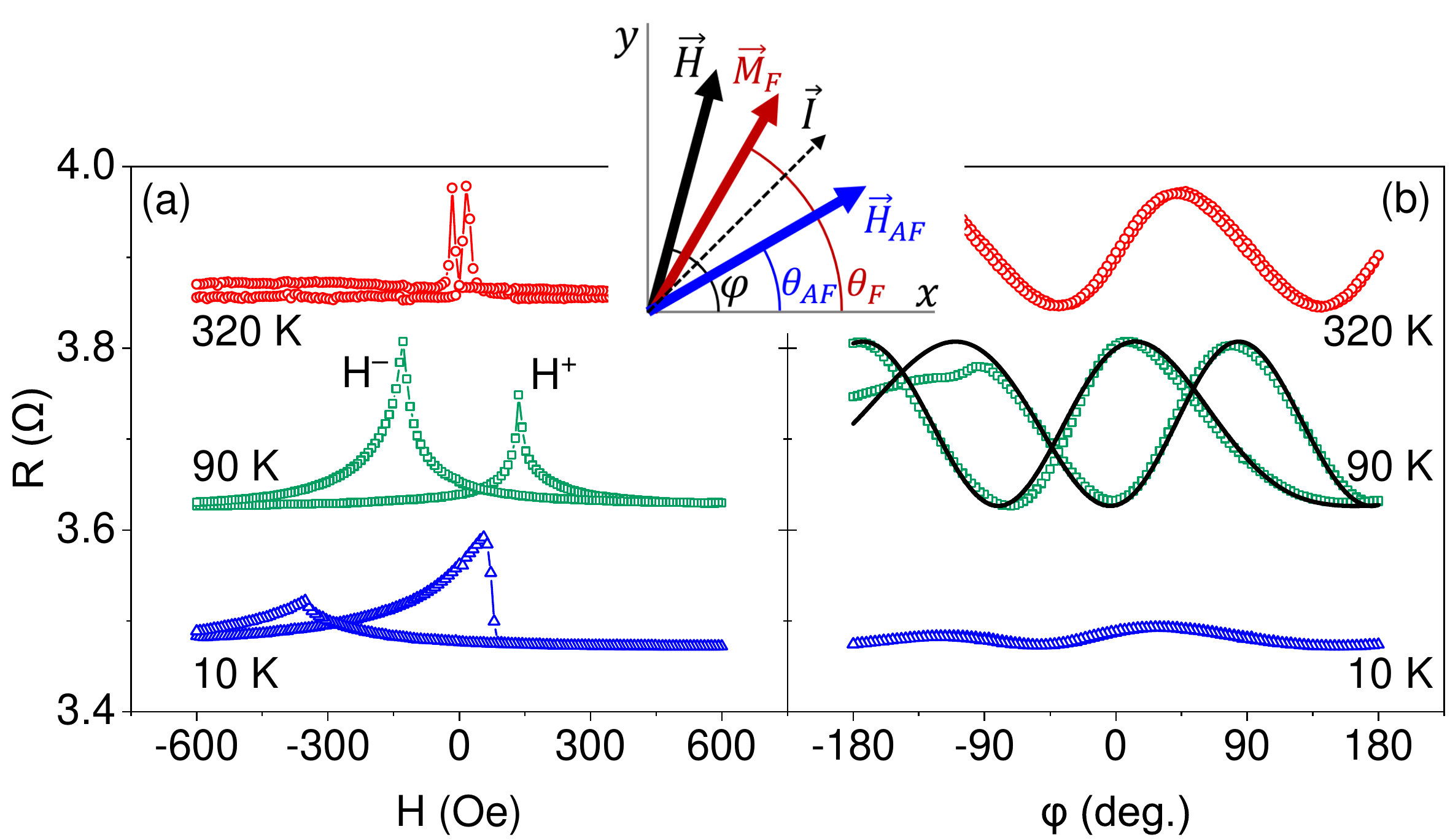}
	\caption{Magnetoelectronic hysteresis loops obtained by varying a field with a fixed $\varphi=-45^\circ$ (a), and by rotating a fixed field $H=180$~Oe at a rate of $3\frac{\text{deg.}}{\text{s}}$ (b), at the labeled values of $T$. The data in (b) are offset for clarity. Curves in (b) are calculations as described in the text, with $\nu = 13\text{ s, } H_{AF} = 155\text{ Oe}$. Inset: schematic of the relevant angles.}
	\label{fig_2}
\end{figure}

{\it Model.} To facilitate the analysis of memristive behaviors in F/AF heterostructures, we introduce a model for the viscous dynamics of AF coupled to F. In our experiment, the resistive signal is provided by the anisotropic magnetoresistance (AMR) - the $180^\circ$-periodic dependence of $R$ on the angle between the electric current $\vec{I}$ and the magnetization $\vec{M}_F$ of F. The relevant angles, measured relative to the fixed x-axis, are introduced in the inset in Fig.~\ref{fig_2}(b): $\varphi$ is the angle of the magnetic field $\vec{H}$, $\theta_F$($\theta_{AF}$) is the angle of $\vec{M}_F$($\vec{M}_{AF}$). Here, $\vec{M}_{AF}$ is the uncompensated magnetization of AF at the interface with F, producing effective exchange field $\vec{H}_{AF}\parallel\vec{M}_{AF}$. The current $\vec{I}$ is directed at a fixed angle of $45^\circ$.

The AMR can be linearized around $\theta_F=0$,
\begin{equation}
    R(\theta_F)=R_0+\Delta R\sin^2(\theta_F+\frac{\pi}{4})\approx R_0 + \frac{\Delta R}{2} + \Delta R \theta_F,
\end{equation}
where $R_0$ is the AMR minimum, and $\Delta R$ is the magnetoresistance.

On timescales significantly larger than that of intrinsic dynamics of F, the direction of $\vec{M}_F$ is determined by the balance between the torques exerted by $\vec{H}$ and $\vec{H}_{AF}$~\cite{URAZHDIN201975},
\begin{equation}\label{eq:torques}
	\vec{H} \times \vec{M}_F = \vec{M}_F \times \vec{H}_{AF}.
\end{equation}

This equation can be rewritten in terms of the relevant angles as
\begin{equation}\label{eq:F}
	H(t)\sin(\varphi(t)-\theta_{F}(t)) = H_{AF}\sin(\theta_F(t)-\theta_{AF}(t)).
\end{equation}

For AF, we assume viscous dynamics described by the magnetic viscosity constant $\nu$ and driven by the exchange interaction with F, 
\begin{equation}\label{eq:AF}
	\nu \frac{\partial \theta_{AF}(t)}{\partial t} = \sin(\theta_F(t)-\theta_{AF}(t)).
\end{equation}

The dynamics of the system involves coupled evolution of F and AF described by Eqs.~(\ref{eq:F}) and (\ref{eq:AF}).
For concreteness, assume that the system is initially prepared in the state $\theta_F=\theta_{AF}=0$, and a time-dependent  magnetic field $H(t)$ is applied at angle $\varphi=90^\circ$ starting at $t=0$. To the lowest order in $\theta_{AF}$, the solution of Eq.~(\ref{eq:AF}) is 
\begin{equation}\label{eq:thetaAF}
\theta_{AF}(t)\approx\frac{1}{\nu H_{AF}}\int_{0}^{t} H(t')dt'.
\end{equation}

For the sinusoidally varying driving magnetic field $H(t)=H_0\sin\frac{2\pi t}{\tau}$,
\begin{equation}\label{eq:thetaAF2}
	\theta_{AF}(t) \approx \frac{\tau H_0}{\pi \nu H_{AF}}(1-\cos\frac{2\pi t}{\tau})
\end{equation}
resulting in an elliptic R-H hysteresis loop with an area
\begin{equation}
	S \approx \Delta R\frac{\tau H_0^2}{\nu H_{AF}},
\end{equation}
proportional to the period $\tau$ of the input and inversely proportional to the magnetic viscosity. A similar result is obtained for the triangular driving signal, establishing the connection with the ideal memristor, see Fig.~\ref{fig_1}(b).

{\it Methods.} The studied sample with the structure  NiO(8~nm)/Py(10~nm)/Ta(5~nm) was deposited on an oxidized Si substrate by high-vacuum magnetron sputtering at room temperature. NiO was deposited by reactive sputtering from a Ni target in Ar/O$_2$ mixture, with the partial oxygen pressure adjusted to optimize the magnetic properties, as in our previous studies of NiO and the structurally similar CoO~\cite{PhysRevB.94.024422,PhysRevB.97.054402,URAZHDIN201975,PhysRevB.101.144427}. Py and Ta were deposited in ultrahigh purity Ar. The Ta capping layer was used to protect the structure from oxidation. The deposition was performed in $\approx 100$ Oe in-plane field known to facilitate well-defined exchange anisotropy in F/AF bilayers. 

The thickness of NiO was chosen so that $T_B\approx 90$~K for the studied structure, is well within the experimentally accessible range. Thicker NiO is expected to enable room-temperature device operation. The roughness of magnetic layers, characterized by atomic force microscopy on separate samples, where deposition was terminated at the respective layer, was $\le 0.35$~nm rms.

Magnetoelectronic measurements were performed in the 4-probe van der Pauw geometry, using ac electric current $I=0.1\text{ mA}$ rms at frequency $f=1.3\text{ kHz}$, and lock-in detection. To initialize the system in a well-defined magnetic state, it was cooled in a constant magnetic field $H=400\text{ Oe}$ oriented at $\varphi=0$ (see inset in Fig.~\ref{fig_2}). The magnetic properties of the system were characterized by magnetoelectronic measurements of the hysteresis loop obtained by sweeping the field at $\varphi=-45^{\circ}$ corresponding to the minimum of AMR (Fig.~\ref{fig_2}(a)), and by the response to rotating field of fixed magnitude (Fig.~\ref{fig_2}(b)).

{\it Characterization.}
The peaks in the hysteresis loops (Fig.~\ref{fig_2}(a)) at fields $H^+$ and $H^-$, are associated with the reversal of $M_F$, allowing us to determine the coercivity $H_C=\frac{H^+ - H^-}{2}$ and the effective EB field $H_E=-\frac{H^+ + H^-}{2}$  (Fig.~\ref{fig_2}(a)). At the temperature $T=320$~K$\gg T_B=90$~K, $H_E=0$ and the coercivity is small. The coercivity is enhanced at $T=90\text{ K}=T_B$, but $H^+$ and $H^-$ remain symmetric. The asymmetry of reversal points at $T=10~{\text{K}}<T_B$ is associated with the onset of EB, resulting in a finite $H_E\approx 160 \text{ Oe}$. These behaviors are consistent with prior studies of F/AF bilayers based on NiO and other AF materials~\cite{Radu2008,BERKOWITZ1999552,NOGUES1999203}.

Figure~\ref{fig_2}(b) shows the response to a rotating field $H=180$~Oe. At $T=320$~K, the dependence $R(\varphi)$ is sinusoidal and the hysteresis with respect to the direction of rotation is negligible, as expected for $M_F$ aligned with the rotating field. At $T=90\text{ K}=T_B$, the dependence remains sinusoidal, consistent with the absence of directional anisotropy, but a large hysteresis appears with respect to the rotation direction. Similar hysteretic behaviors have been reported at $T_B$ for other F/AF systems and interpreted in terms of rotating exchange anisotropy~\cite{PhysRevB.70.094420,PhysRevB.71.220410}. At $T=10$~K, the resistance remains almost constant, indicating that $M_F$ is pinned by EB.

The hysteretic response to the rotating field at $T=90\text{ K}$ shown in Fig.~\ref{fig_2}(b) indicates that the state of the magnetic system is history-dependent, as required for the memristive functionality. In particular, these data show that the value of $R$ at $\varphi=0$ can vary over almost the entire range of AMR, i.e. the metastable orientations of $\vec{M}_F$ span the range of almost $180^\circ$ even at a relatively large $H=180$~Oe oriented at $\varphi=0$. These data are well described by the model of viscous AF magnetization dynamics introduced above, as shown by the curves in Fig.~\ref{fig_2}(b) obtained by solving Eqs.~(\ref{eq:F}), (\ref{eq:AF}). 

{\it Memristive functionality.} We now experimentally demonstrate that the studied structure exhibits nearly ideal memristive functionality at temperatures above $T_B=90$~K, and confirm its origin from viscous AF magnetization dynamics.

\begin{figure}
	\includegraphics[width=\columnwidth]{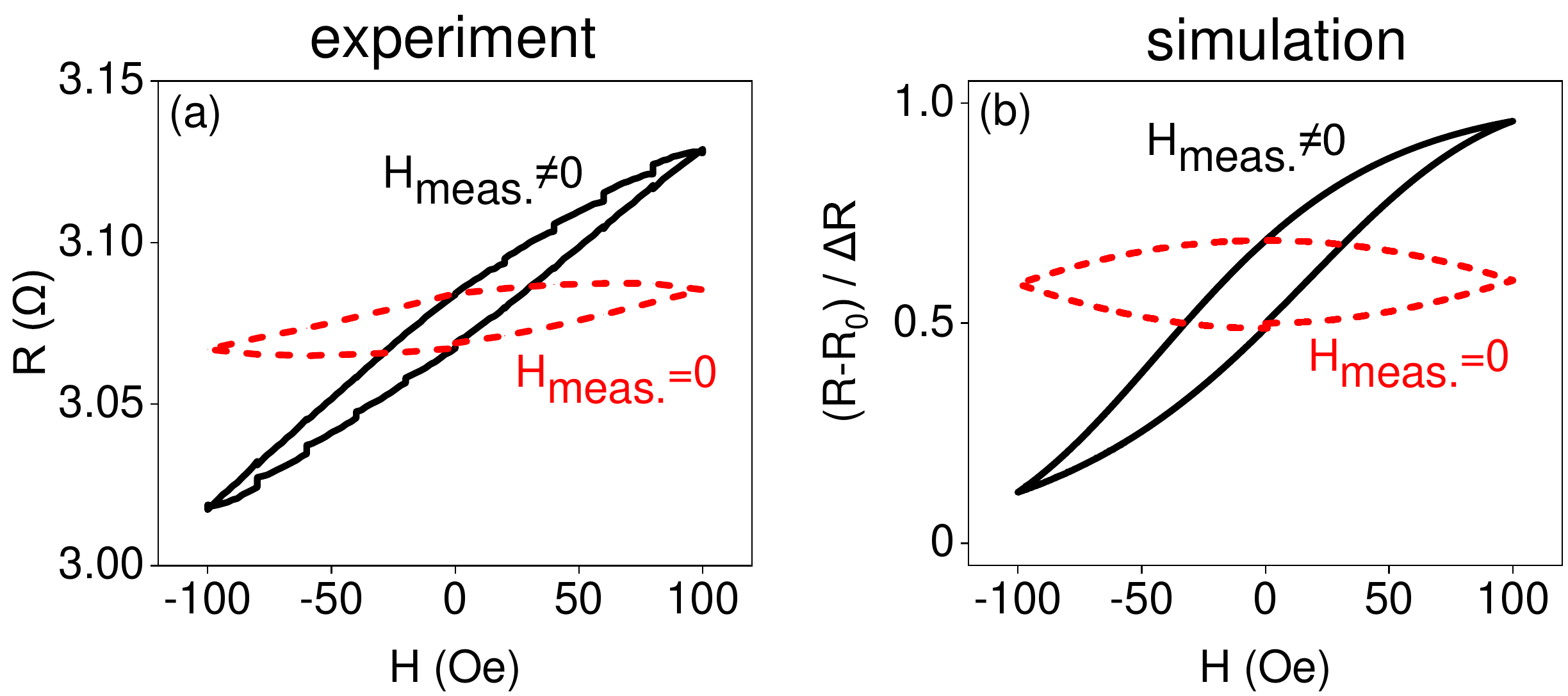}
	\caption{(a) Simulated $R(\theta_{F})$ vs $H$ (solid curve) and $R(\theta_{AF})$ vs $H$ (dashed curve). (b) $R(\theta_{F})$ (solid curve) and $R(\theta_{AF})$ (dashed curve) measured at $T=90$~K. The plotted quantities are defined in the text.}
	\label{fig_3}
\end{figure}

To test the memristive functionality, the system is initialized in a state with $\theta_F=\theta_{AF}=0$, corresponding to the middle of the linear part of AMR, by in-field cooling. Magnetic field perpendicular to the magnetization ($\varphi=90^{\circ}$) is subsequently periodically swept in a triangular wave pattern, and the sample resistance is simultaneously measured.

Solid curve in Fig.~\ref{fig_3}(a) shows the $R$ vs $H$ hysteresis loop obtained at $T=90$~K by sweeping $H$ between $-100$~Oe and $100$~Oe with a period of $100$~s. Its shape is reminiscent of the lens-shaped hysteresis curve expected for an ideal memristor (Fig.~\ref{fig_1}(b)), but is tilted. The tilt can be explained by the hybrid nature of the magnetic system, where the AF providing memristive functionality is coupled only indirectly to the driving field via F. 

To directly access the state of AF, we modified the measurement protocol to remove the external field prior to each measurement during the sweep. Since the intrinsic dynamics of F occurs on significantly shorter timescales than our measurement, $M_F$ aligns with $M_{AF}$, so the state of AF is revealed by such measurements. This modified approach almost completely eliminated the tilt of the hysteresis loop (dashed curve in Fig.~\ref{fig_3}(a)), confirming its origin from the hybrid nature of the magnetic system. Simulations based on the model of viscous dynamics introduced above reproduces the salient features of these results (Fig.~\ref{fig_3}(b)).

The hysteresis in Fig.~\ref{fig_3} demonstrates generalized memristive functionality - dependence of resistance on the history of the applied magnetic field. To investigate its relationship to the ideal memristor, we analyze the dependence of the area under the hysteresis loop on the period of the field sweep. An ideal memristor is expected to exhibit a linear dependence of $R$ on the period of the driving input, as discussed above. 

\begin{figure}[ht]
	\includegraphics[width=\columnwidth]{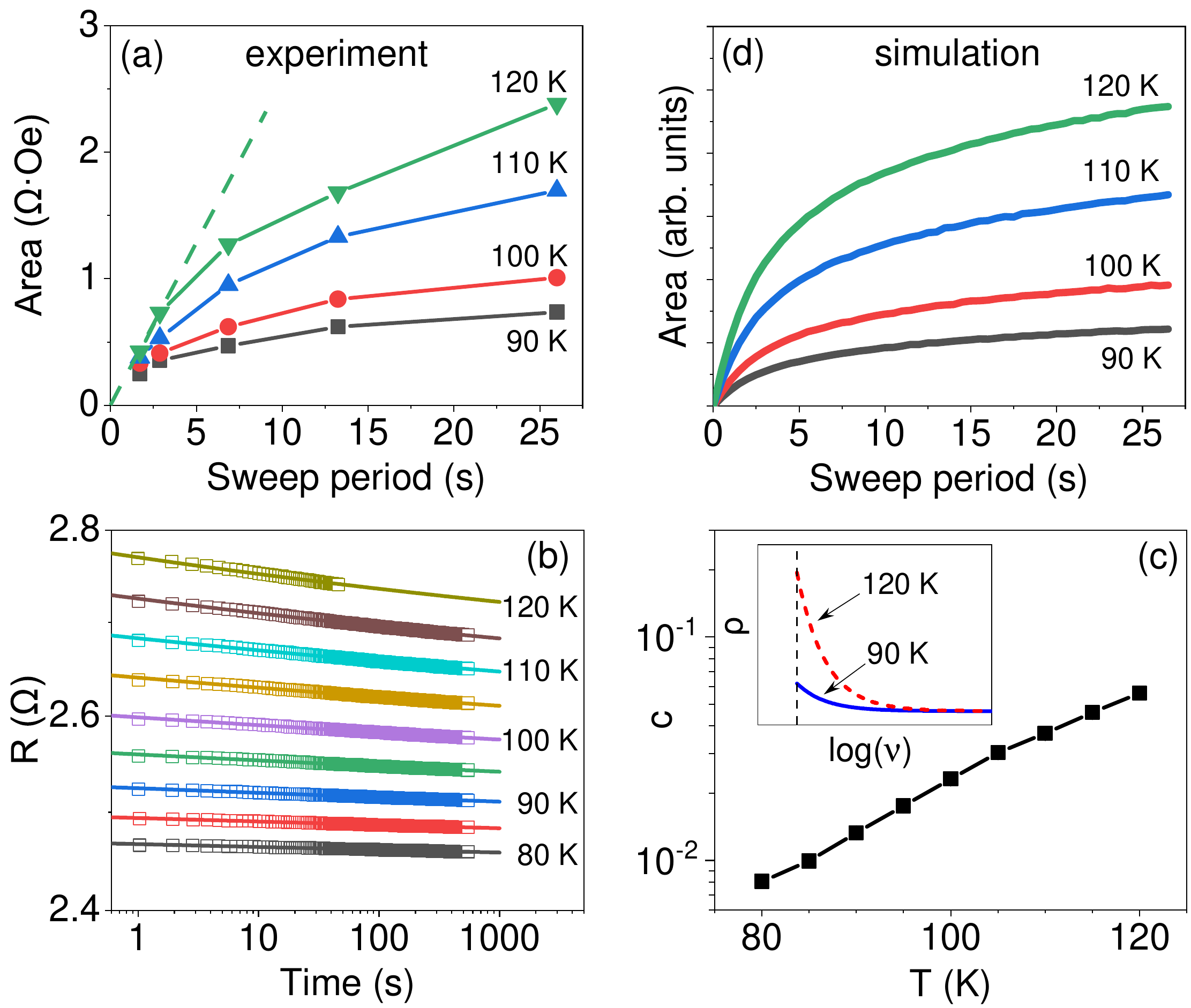}
	\caption{(a) Measured dependence of the area of the $R$ vs $H$ hysteresis loop on the period of the applied field sweep, at the labeled values of $T$.  (b) Symbols: $R$ vs time recorded after abruptly rotating the field $H_0=100$~Oe from $\varphi=45^{\circ}$ to $\varphi=-45^{\circ}$, at the labeled values of $T$. Curves: fitting with $ R(t) = R(\infty) + At^{-c} $. (c) Dependence of the power-law exponent $c$ on $T$. Inset: distribution of viscosity extracted from the aging data, for two values of $T$. (d) Dependence of the area under the $R-H$ hysteresis loop on the period of the input signal, simulated using the magnetic viscosity distribution determined from the aging data.}
	\label{fig_4}
\end{figure}

Figure~\ref{fig_4}(a) shows the measured dependences of the area of the $R$ vs $H$ hysteresis loop on the period of the input, for $T=90-120$~K. These data show that the measured area increases with the period of the input, as expected for the ideal memristor, but the dependence is nonlinear. The area also increases with increasing $T$, consistent with the decrease of magnetic viscosity.

We now show that the deviations from the ideal memristive behaviors manifested by the nonlinear dependences in Fig.~\ref{fig_4}(a) can be explained by the inhomogeneity of viscous AF dynamics, resulting in a distribution of viscosity $\nu$. Such inhomogeneity may result from the local variations of the AF thickness and/or from the inherent local variations of the effects of frustration due to random exchange at F/AF interface.

To determine the distribution $\rho(\nu)$ of the magnetic viscosity, we performed separate measurements of magnetic aging, in which the magnetic field was abruptly rotated, and subsequently the time-dependent resistance of the sample was monitored at a fixed field $H_0=100$~Oe. For a system characterized by a single value of viscosity $\nu$, the dependence predicted by our model is
\begin{equation}
	\label{eq:11}
	R(t) - R(\infty) \approx \Delta R \left(\frac{H_{AF}^2}{H_{AF}^2+H_0^2} \right) e^{-\frac{2H_0}{\nu(H_0+H_{AF})}t},
\end{equation}
where $R(\infty)$ is the asymptotic value of $R$ corresponding to $\vec{M}_F\parallel \vec{H}_0 $.

As shown in Fig.~\ref{fig_4}(b), aging does not follow the exponential dependence of Eq.~(\ref{eq:11}), but can be well-fitted by the power-law dependence on time, in the form $ R(t) = R(\infty) + At^{-c} $, where $A$ and $c$ are constants. Power-law dependence indicates that there is no characteristic time scale in the system, i.e. no single viscosity value, but a distribution of viscosity throughout the sample. An alternative, or a co-existing, possible mechanism is a broad distribution of the effective exchange field $H_{AF}$, which enters Eq.~(\ref{eq:11}) on equal footing with $\nu$.

We note that the observation of aging provides definitive proof for viscous dynamics, as opposed to dry friction~\cite{PhysRevApplied.12.044029,9108129}. Indeed, once the dry friction is overcome at sufficiently large $H_0$, the magnetic system would rapidly evolve and relax over the intrinsic dynamical scales of order nanosecond. The measured power-law aging can be interpreted as a superposition of contributions from the exponential aging associated with a particular value of $\nu$, and characterized by the distribution $\rho(\nu)$~\cite{PhysRevB.94.024422},
\begin{equation}
R(t) - R(\infty) = At^{-c} \propto \int_{0}^{\infty} \rho(\nu) e^{-\frac{2H_0}{\nu(H_0+H_{AF})}t} d\nu.
\end{equation}

The distribution should be normalized
\begin{equation}
\int_{\nu_{min}}^{\infty} \rho(\nu) d\nu = 1,
\end{equation}
where $\nu_{min}$ is the cutoff. Physically, at short timescales corresponding to small $\nu$, the system must crossover from viscous to damped dynamics, and the dependence Eq.~(\ref{eq:11}) becomes inapplicable. We use the value  $\nu_{min}=0.1$~s which is below the shortest timescales in our measurements, and does not influence the quality of fitting.

Inverse Laplace transform yields Pareto distribution for the viscosity,
\begin{equation}\label{eq:Pareto}
\rho(\nu) = \frac{c\nu^{-(c+1)}}{\nu_{min}^{-c}}.
\end{equation}
The distribution extracted from the data is shown in the inset of Fig.~\ref{fig_4}c for two values of $T$.
The power-law exponent $c$ rapidly increases with the temperature (Fig.~\ref{fig_4}(c)). According to Eq.~(\ref{eq:Pareto}), this implies a redistribution of $\rho(\nu)$ towards smaller values of $\nu$, consistent with the expected decrease of viscosity.

Figure~\ref{fig_4}(d) shows the simulated dependence of the R-H hysteresis loop area on the period of the field sweep and temperature, obtained using the distribution of viscosity determined from the aging data. These results reproduce the observed deviations from the ideal memristive behaviors (Fig.~\ref{fig_4}(a)), supporting our interpretation of these deviations in terms of the inhomogeneous distribution of viscosity.

\section{Summary}
We have experimentally demonstrated memristive behaviors in a thin-film F/AF bilayer, where  frustration at the F/AF interface leads to viscous dynamics of magnetization. 
We showed that in contrast to the resistive switching memory or dry friction-based magnetic memristors, viscous dynamics results in the integral dependence of device resistance on its input, consistent with the original definition of an ideal memristor. The demonstrated devices exhibit deviations from the ideal  behaviors, due to the distribution of viscosity associated with the frustrated nature of the studied system and/or the non-negligible effects of film thickness variations.

In our proof-of-principle demonstration, external magnetic field was utilized as the input driving the magnetization dynamics, while anisotropic magnetoresistance was utilized as the mechanism to convert these dynamics into resistance variations. For applications, the proposed structures must be downscaled to nanometer dimensions, anisotropic magnetoresistance should be replaced with tunneling magnetoresistance to provide much larger signals, and the devices should be driven by spin torque instead of the external magnetic field. An important issue related to the practicality of the proposed memristors is whether the magnetic frustration underlying viscous dynamics will result in large device-to-device variations at the nanoscale. While such crossover must inevitably occur at sufficiently small lengthscales, a prior observation that aging in similar F/AF heterostructures is associated with collective dynamics~\cite{PhysRevB.92.174416,PhysRevB.94.024422} is encouraging, since it implies that the effects of individual magnetic defects and pinning centers are averaged by internal magnetic interactions. Along with the well-known high endurance of magnetic devices, the demonstrated nearly ideal memristive behaviors should make the proposed devices based on viscous magnetization dynamics well-suited for the hardware implementation of artificial neural networks.

\section{Acknowledgments}
This work was supported by NSF awards ECCS-1804198 and ECCS-2005786.

The data that support the findings of this study are available from the corresponding author upon reasonable request.

\bibliography{mybib}
\bibliographystyle{apsrev4-1}
\end{document}